\PassOptionsToPackage{unicode}{hyperref}
\PassOptionsToPackage{hyphens}{url}
\PassOptionsToPackage{dvipsnames,svgnames,x11names}{xcolor}
\documentclass[
  12pt]{article}
\usepackage{mathrsfs}
\usepackage{comment}
\usepackage{amsmath,amssymb}
\usepackage{iftex}
\ifPDFTeX
  \usepackage[T1]{fontenc}
  \usepackage[utf8]{inputenc}
  \usepackage{textcomp} 
\else 
  \usepackage{unicode-math}
  \defaultfontfeatures{Scale=MatchLowercase}
  \defaultfontfeatures[\rmfamily]{Ligatures=TeX,Scale=1}
\fi
\usepackage{lmodern}
\ifPDFTeX\else
\fi
\IfFileExists{upquote.sty}{\usepackage{upquote}}{}
\IfFileExists{microtype.sty}{
  \usepackage[]{microtype}
  \UseMicrotypeSet[protrusion]{basicmath} 
}{}
\makeatletter
\@ifundefined{KOMAClassName}{
  \IfFileExists{parskip.sty}{%
    \usepackage{parskip}
  }{
    \setlength{\parindent}{0pt}
    \setlength{\parskip}{6pt plus 2pt minus 1pt}}
}{
  \KOMAoptions{parskip=half}}
\makeatother
\usepackage{xcolor}
\makeatletter
\ifx\paragraph\undefined\else
  \let\oldparagraph\paragraph
  \renewcommand{\paragraph}{
    \@ifstar
      \xxxParagraphStar
      \xxxParagraphNoStar
  }
  \newcommand{\xxxParagraphStar}[1]{\oldparagraph*{#1}\mbox{}}
  \newcommand{\xxxParagraphNoStar}[1]{\oldparagraph{#1}\mbox{}}
\fi
\ifx\subparagraph\undefined\else
  \let\oldsubparagraph\subparagraph
  \renewcommand{\subparagraph}{
    \@ifstar
      \xxxSubParagraphStar
      \xxxSubParagraphNoStar
  }
  \newcommand{\xxxSubParagraphStar}[1]{\oldsubparagraph*{#1}\mbox{}}
  \newcommand{\xxxSubParagraphNoStar}[1]{\oldsubparagraph{#1}\mbox{}}
\fi
\makeatother

\usepackage{longtable,booktabs,array}
\usepackage{calc} 
\usepackage{etoolbox}
\makeatletter
\patchcmd\longtable{\par}{\if@noskipsec\mbox{}\fi\par}{}{}
\makeatother
\IfFileExists{footnotehyper.sty}{\usepackage{footnotehyper}}{\usepackage{footnote}}
\makesavenoteenv{longtable}
\usepackage{graphicx}
\makeatletter
\def\maxwidth{\ifdim\Gin@nat@width>\linewidth\linewidth\else\Gin@nat@width\fi}
\def\maxheight{\ifdim\Gin@nat@height>\textheight\textheight\else\Gin@nat@height\fi}
\makeatother
\setkeys{Gin}{width=\maxwidth,height=\maxheight,keepaspectratio}
\makeatletter
\def\fps@figure{htbp}
\makeatother

\makeatletter
\@ifpackageloaded{caption}{}{\usepackage{caption}}
\AtBeginDocument{%
\ifdefined\contentsname
  \renewcommand*\contentsname{Table of contents}
\else
  \newcommand\contentsname{Table of contents}
\fi
\ifdefined\listfigurename
  \renewcommand*\listfigurename{List of Figures}
\else
  \newcommand\listfigurename{List of Figures}
\fi
\ifdefined\listtablename
  \renewcommand*\listtablename{List of Tables}
\else
  \newcommand\listtablename{List of Tables}
\fi
\ifdefined\figurename
  \renewcommand*\figurename{Figure}
\else
  \newcommand\figurename{Figure}
\fi
\ifdefined\tablename
  \renewcommand*\tablename{Table}
\else
  \newcommand\tablename{Table}
\fi
}
\@ifpackageloaded{float}{}{\usepackage{float}}
\floatstyle{ruled}
\@ifundefined{c@chapter}{\newfloat{codelisting}{h}{lop}}{\newfloat{codelisting}{h}{lop}[chapter]}
\floatname{codelisting}{Listing}

\makeatother
\makeatletter
\@ifpackageloaded{caption}{}{\usepackage{caption}}
\@ifpackageloaded{subcaption}{}{\usepackage{subcaption}}
\makeatother

\ifLuaTeX
  \usepackage{selnolig}  
\fi
\usepackage[]{natbib}
\usepackage{bookmark}
\usepackage{graphicx}
\IfFileExists{xurl.sty}{\usepackage{xurl}}{} 
\hypersetup{
  pdftitle={Title},
  pdfauthor={Author 1; Author 2},
  pdfkeywords={3 to 6 keywords, that do not appear in the title},
  colorlinks=true,
  linkcolor={blue},
  filecolor={Maroon},
  citecolor={Blue},
  urlcolor={Blue},
  pdfcreator={LaTeX via pandoc}}

\newcommand{\anon}{1}

\begin{document}

\def\spacingset#1{\renewcommand{\baselinestretch}%
{#1}\small\normalsize} \spacingset{1}


\if1\anon
{
  \title{\bf Consistent and powerful CUSUM change-point test for panel data with changes in variance}
  \author{Wenzhi Yang \hspace{.2cm}\\
    School of Big Data and Statistics, Anhui University, Hefei, China\\
    Yueting Xu \\
    School of Big Data and Statistics, Anhui University, Hefei, China\\
        Xiaoping Shi\\
    Irving K. Barber Faculty of Science, \\University of British Columbia, Kelowna, Canada\\
    and \\
        Qiong Li \thanks{Corresponding author: qiongli@bnbu.edu.cn
    }\\
    Guangdong Provincial Key Laboratory of \\Interdisciplinary Research and Application for Data \\Science,  Beijing Normal-Hong Kong Baptist University, Zhuhai, China}
  \maketitle
} \fi

\if0\anon
{
  \bigskip
  \bigskip
  \bigskip
  \begin{center}
    {\LARGE\bf Title}
\end{center}
  \medskip
} \fi

\bigskip
\begin{abstract}
This paper investigates the change-point of variance in panel data models with time series of $\alpha$-mixing. Based on the cumulative sum (CUSUM) method and the individual differences, we construct a CUSUM test for panel data models to detect variance changes. Under the null hypothesis, we derive the limit distribution of this test, which can be used to detect the change-point of variance.
Under the alternative hypothesis, the limit behavior of the CUSUM test is also derived. To validate the performance of the test, we conducted simulation analyses with Gaussian and Gamma errors. The results demonstrate that this testing method significantly outperforms existing approaches, particularly in detecting sparse variance changes. Finally, we conducted a practical case study using panel data from the Shanghai Shenzhen CSI 300 Index Components. Not only did we successfully identify the change-points of variance, but we also delved deeper into the underlying economic drivers behind these changes.
\end{abstract}

\noindent%


\section{Introduction}

Change-point detection is a statistical methodology concerned with identifying points in time (or sequence order) at which the underlying probability distribution of a data-generating process undergoes a significant alteration. These points, often referred to as change-points, breakpoints, or structural breaks, represent moments of regime shifts, transitions, or discontinuities in the system being observed \citep{Csorgo1997, Chen2012}. Panel data, comprising observations on multiple entities (individuals, firms, countries, households, etc.) over multiple time periods, combine both cross-sectional and time-series dimensions, enabling more powerful detection of structural changes, including both mean shifts and variance shifts.

Detecting variance changes in panel data is particularly important in applications such as finance, genomics, and sensor networks, where volatility shifts may occur in only a subset of units. For instance, in financial markets, occasional shocks may affect only a few assets; in genomics, mutations may induce variance changes in some genes but not others; and in sensor networks, sporadic sensor failures or anomalies can lead to localized variance changes. Classical approaches, such as the test of \cite{li2015}, aggregate panel-wise CUSUM statistics of squared residuals normalized by individual long-run variances. While asymptotically valid, these methods can lose power under sparse alternatives, where only a few panels undergo variance changes.

To address this issue, we propose a new statistic, which aggregates squared residuals across panels prior to normalization, preserving signal strength in affected panels. The panel errors are modeled as stationary $\alpha$-mixing sequences, allowing for weak temporal dependence within each panel while maintaining independence across panels. Its performance is evaluated using two complementary criteria: (i) the signal-to-noise ratio (SNR), which quantifies the strength of the signal relative to the variability under the alternative hypothesis, and (ii) stochastic comparison, defined as the probability that one statistic exceeds another under the alternative. Monte Carlo simulations demonstrate the superior power of our new statistic in detecting sparse variance changes; see Section~2 for details.

There has been a growing literature on change-point detection in panel data. For example, \cite{Bai2010} studied common breaks in means and variances using least squares and quasi-maximum likelihood; \cite{horvan2012} proposed a centered CUSUM statistic for mean changes; \cite{li2015} extended this framework to variance changes; \cite{Cho2016} proposed a double CUSUM approach for constant signals; \cite{Xu2016} introduced weighted differences of averages for dependent panels; \cite{Choi2021} gave a general panel break test based on the self-normalization method; and several recent works \citep{Jin2016, Shi2017, Avanesov2018, Wang2018, Liu2021, li2023, Duker2024, Shao2010, She2025, Lumsdaine2023, Wang2024, Loyo2025} have addressed high-dimensional and sparse change scenarios. In addition, alternative variance change-point tests have been proposed, including the ICSS algorithm \citep{inclan}, wavelet-based methods \citep{wavmethod2005}, and likelihood-ratio tests. These approaches often assume independence, focus on univariate series, or are sensitive to noise. By contrast, our CUSUM-based modification is robust to weak dependence, accommodates irregular panels, and directly addresses sparse variance changes, making it well-suited for practical applications where only a subset of panels experiences variance shifts.

The rest of this paper is organized as follows. Section 2 introduces the panel variance change-point model and the construction of the proposed test. Section 3 presents the theoretical asymptotic analysis, including the distribution under the null hypothesis and limit behavior under the alternative. Section 4 illustrates the test via Monte Carlo simulations for Gaussian and Gamma error panels, highlighting superior performance compared to the test of \cite{li2015}. Section 5 provides a real data application to the Shanghai Shenzhen CSI 300 Index Components to detect variance change-points. Finally, Section 6 concludes, and Appendix A contains the proofs.

\section{Panel Variance Change-Point Models and Test Construction}
\subsection{Variance Change-Point Test of Li et al.\ (2015)}

Detecting variance changes is particularly important in econometrics, finance, genomics, and sensor networks, where volatility or dispersion may shift in a subset of cross-sectional units. \cite{li2015} studied the problem of detecting a variance change-point in panel data models.
They considered the panel data model
\begin{equation}
X_{i,t} = \mu_i + e_{i,t},
\quad 1 \le i \le N,\; 1 \le t \le T,
\label{panel_li}
\end{equation}
where $\mu_i$ denotes an individual-specific mean
and $\{e_{i,t}\}$ are zero-mean error sequences of linear processes which are assumed to be weakly dependent over time and independent across panels.
The error variance is allowed to change at an unknown common time point $t^* \in \{1,\ldots, T-1\}$, such that
\begin{equation}
\lim_{T\rightarrow\infty}\frac{1}{T}\operatorname{Var}\Big(\sum_{t=1}^Te_{i,t}\Big) =
\begin{cases}
\sigma_i^2, & t \le t^*, \\
\sigma_i^2 + \delta_i, & t > t^*,
\end{cases}
\quad 1 \le i \le N.
\label{var_li}
\end{equation}

The null hypothesis of no variance change is given by
\begin{equation}
H_0:\quad \delta_1 = \delta_2 = \cdots = \delta_N = 0,
\label{H0_li}
\end{equation}
while the alternative hypothesis assumes a common variance increase across all panels,
\begin{equation}
H_1:\quad \delta_i > 0 \quad \text{for all } i = 1,\ldots,N.
\label{H1_li}
\end{equation}
\cite{li2015} constructed a centered CUSUM statistic of squared residuals,
\begin{equation}
C_{i,T}(k)
=
\frac{1}{\sqrt{T}}
\left(
\sum_{t=1}^k \hat{e}_{i,t}^2
-
\frac{k}{T}\sum_{t=1}^T \hat{e}_{i,t}^2
\right),
\quad 1 \le k \le T-1,
\label{CUSUM_i}
\end{equation}
where $\hat{e}_{i,t}=X_{i,t}-\overline{X}_i$ and $\overline{X}_i = \frac{1}{T}\sum_{t=1}^T X_{i,t}$
denotes the sample mean of panel $i$. For each panel. Each panel-wise statistic is normalized by a long-run variance estimator $\hat{s}_{i,T}^{2*}$ of $\lim\limits_{T\to\infty}\frac{1}{T}\operatorname{Var}(\sum_{t=1}^T (X_{i,t}-\mu_i)^2)=\lim\limits_{T\to\infty}\frac{1}{T}\operatorname{Var}(\sum_{t=1}^T e_{i,t}^2)$ in \eqref{panel_li}, and the aggregated statistic is defined as
\begin{equation}
V_{N,T}(k)
=
\frac{1}{\sqrt{N}}
\sum_{i=1}^N
\frac{1}{\sqrt{\hat{s}_{i,T}^{2*}}}
\, C_{i,T}(k).
\label{V_stat_li}
\end{equation}
The corresponding test statistic is
\begin{equation}
T_V
=
\max_{1 \le k \le T-1}
\left| V_{N,T}(k) \right|.
\label{TV_li}
\end{equation}

Under the null hypothesis $H_0$ and suitable regularity conditions, \cite{li2015} showed that
\begin{equation}
T_V
\;\xrightarrow{d}\;
\sup_{0 \le x \le 1} |B^0(x)|,
\label{limit_li}
\end{equation}
where $\{B^0(x),\,0\le x\le 1\}$ denotes a standard Brownian bridge. By the (11.38) of \cite{Billingsley}, it shows
that $P(\sup\limits_{0\leq t\leq 1}|B^{0}(t)|\leq x)=1+2\sum\nolimits_{k=1}^\infty
(-1)^{k}e^{-2k^2x^2}$, $x>0$. The critical values of $\sup\limits_{0\leq t\leq 1}|B^{0}(t)|$ are also given in Table 1 of \cite{inclan}, where the 10\%, 5\% and 1\% critical values of $\sup\limits_{0\leq t\leq 1}|B^{0}(t)|$ are 1.224, 1.358 and 1.628.

\subsection{Motivation for a Modified Test}

The variance change-point test of \cite{li2015} provides an important foundation for detecting volatility shifts in panel data. However, both the modeling framework and the construction of their test statistic impose restrictions that can substantially reduce power in empirically relevant situations, especially when variance changes are heterogeneous or sparse across panels. In this subsection, we introduce a more general panel data model and explain the motivation for a modified CUSUM-based test that overcomes these limitations.

\subsubsection{Modeling framework}

Since $\alpha$-mixing \citep{HallandHeyde1980} contains linear and nonlinear processes and it is widely used in the fields of time series, we will study the change-point of variance of the panel data model with $\alpha$-mixing errors. Let us recall the definition of $\alpha$-mixing sequence.

\textit{Definition 2.1 ($\alpha$-mixing sequence)}. Denote $\mathcal{N}=\{1,2,\cdots,n,\cdots\}$. The sequence $\{e_n,n\geq 1\}$ is called strong mixing or $\alpha$-mixing if the $\alpha$-mixing coefficient
\begin{equation}
\alpha(n)=\sup\limits_{m\in \mathcal{N}}\sup\limits_{A\in\mathcal{F}_1^m,B\in\mathcal{F}_{m+n}^\infty}|P(A\cap B)-P(A)P(B)|,\nonumber
\end{equation}
converges to zero as $n\rightarrow\infty$, where $\mathcal{F}_m^n$ denotes the
$\sigma$-field generated by $e_m,e_{m+1},\cdots,e_n$ with $m\leq n$.

Many linear and nonlinear time series models satisfy the mixing properties \citep{HallandHeyde1980}. For example, we consider an infinite order moving average (MA($\infty$)) process
$e_t=\sum\nolimits_{i=0}^\infty a_i\varepsilon_{t-i}, t\geq 1$, where $a_i\to 0$ exponentially fast, and $\{\varepsilon_t\}$ is an $i.i.d.$ sequence. If the probability density function of $e_t$ exists (such as Normal, Cauchy, Exponential), then $\{e_t\}$ is an $\alpha$-mixing with exponentially decaying coefficients. The strictly stationary time series, including the autoregressive moving average (ARMA) processes and geometrically ergodic Markov chains, are the $\alpha$-mixing processes. For more properties and applications, we can refer to \cite{Gyorfi1989, HallandHeyde1980}, etc.

In contrast to \cite{li2015}, who characterize variance changes through shifts in the long-run variance of the error process and implicitly focus on common variance increases across all panels, we consider the following variance change-point panel data model: there exists an unknown time point $t^* \in \{1,\ldots, T-1\}$ such that
\begin{equation}
X_{i,t} = \mu_i + (\sigma_i + \delta_i I(t>t^*)) e_{i,t},
\quad 1 \le i \le N,\; 1 \le t \le T,
\label{a1}
\end{equation}
where $\mu_i$, $\sigma_i$, $\delta_i$, and $t^*$ are unknown parameters. For each $i$, $\{e_{i,t}\}_{t=1}^T$ is a stationary $\alpha$-mixing sequence with $
Ee_{i,t}=0$, $\operatorname{Var}(e_{i,t})=1$ and the error processes are independent across panels.

This formulation \eqref{a1} differs from \eqref{panel_li} by \cite{li2015} in several important aspects. First, both the baseline scale $\sigma_i$ and the magnitude of the variance change $\delta_i$ are allowed to vary across panels, permitting heterogeneous volatility dynamics. Second, variance changes may occur only in a subset of panels, rather than being common to all units. Third, the $\alpha$-mixing assumption on $\{e_{i,t}\}$ substantially weakens the linear process assumption adopted in \cite{li2015}, allowing for a much broader class of temporal dependence structures encountered in practice.

The null and alternative hypotheses are formulated as
\begin{flalign}
H_0:&\quad \delta_1 = \cdots = \delta_N = 0, \label{a2}\\
H_1:&\quad \delta_i \neq 0 \text{ for some } i \in \{1,\ldots,N\}, \text{ at an unknown } t^*. \label{a3}
\end{flalign}
Unlike the alternative in \cite{li2015}, which assumes $\delta_i>0$ for all panels, our alternative explicitly accommodates \emph{sparse variance changes}, where only a small fraction of panels experience a volatility shift.

\subsubsection{Comparison of test construction}
\cite{li2015} constructed a panel statistic by first forming panel-wise CUSUM statistics of squared residuals and then normalizing each panel individually by its long-run variance estimator before aggregation. While this normalization guarantees asymptotic pivotality under the null hypothesis, it can severely attenuate the contribution of panels exhibiting variance changes when such changes are sparse.

To address this issue, we propose a modified CUSUM statistic that aggregates information across panels \emph{prior} to normalization:
\begin{equation}
U_{N,T}(k)
=
\frac{k(T-k)}{T}
\Big(
\frac{1}{k}\sum_{t=1}^k \sum_{i=1}^N \hat{e}_{i,t}^2
-
\frac{1}{T-k}\sum_{t=k+1}^T \sum_{i=1}^N \hat{e}_{i,t}^2
\Big), 1\leq k\leq T-1,\label{a4}
\end{equation}
where $\hat{e}_{i,t}=X_{i,t}-\overline{X}_i$ and $\overline{X}_i=\frac{1}{T}\sum\nolimits_{t=1}^T X_{it}$.
The resulting test statistic is
\begin{equation}
T_U
=
\frac{1}{\sqrt{T \sum_{i=1}^N \hat{s}_{i,T}^{2*}}}
\max_{1 \le k \le T-1}
|U_{N,T}(k)|
\;\xrightarrow{d}\;
\sup_{0 \le x \le 1} |B^0(x)|,
\label{TU_stat}
\end{equation}
where $\hat{s}_{i,T}^{2*}$ is a long-run variance estimator of $\sigma_i^4 \lim\limits_{T\to\infty}\frac{1}{T}\operatorname{Var}(\sum_{t=1}^Te_{i,t}^2)$ in \eqref{a1}.

At the significance level $\alpha=0.05$, if $T_{U}$ defined by \eqref{TU_stat} is bigger than 1.358 (or the $p$-value of $T_U$ defined by $P(\sup\limits_{0\leq t\leq 1}|B^{0}(t)|> T_U)$  is smaller than 0.05), then we reject $H_0$: $\delta_1=\delta_2=\ldots=\delta_N=0$ and make a conclusion that there is a change-point of variance in the panel data model \eqref{a1}, and the change-point time location $t^*$ is suggested by
\begin{equation}
\hat{t}_{N,T}=\mathop{\arg \max}\limits_{1\leq k\leq  T-1}\frac{1}{\sqrt{T\sum_{i=1}^N\hat{s}_{i,T}^{2*}}}
|U_{N,T}(k)|=\mathop{\arg \max}\limits_{1\leq k\leq  T-1}|U_{N,T}(k)|.\label{cr1}
\end{equation}

\subsubsection{Power comparison}

The principal advantage of the proposed statistic $T_U$ lies in its superior power under sparse variance-change alternatives. In this subsection, we compare the power of $T_U$ with that of the test statistic $T_V$ proposed by \cite{li2015} using two complementary criteria: (i) signal-to-noise ratio analysis and (ii) stochastic comparison under the alternative hypothesis.

\textit{Definition 2.2 (Signal-to-noise ratio).}
Let $T$ be a test statistic that can be decomposed under the alternative hypothesis $H_1$ as
$T = \text{signal} + \text{noise}$.
The signal-to-noise ratio (SNR) of $T$ is defined as
\[
\text{SNR}(T) = \frac{E(T \mid H_1)}{\sqrt{\operatorname{Var}(T \mid H_1)}}.
\]

A larger SNR indicates stronger separation between the null and alternative distributions and hence higher detection power.

\textit{Definition 2.3 (Stochastic comparison).}
  We say that $T_U$ is more powerful than $T_V$ under the alternative hypothesis $H_1$ if
\[
P(T_U > T_V \mid H_1) > \frac{1}{2},
\]
that is, $T_U$ stochastically dominates $T_V$ under $H_1$.

Although $T_U$ and $T_V$ share the same asymptotic distribution under $H_0$, under $H_1$ the stochastic comparison shows that $T_U$ tends to take larger values than $T_V$, reflecting its higher detection power for sparse variance changes.

We illustrate the power advantage of the proposed statistic under sparse variance changes using a simple simulation.
We consider a panel data model with a single common variance change-point,
$X_{i,t} = \mu_i+(\sigma_{i}+\delta_iI(t>t^*)) e_{i,t}$, $1 \le i \le N$, $1 \le t \le T$, where $t^*$ denotes the common change-point and $\{e_{i,t}\}$ are independent and identically distributed ($i.i.d.$ standard normal random variables. Let $\mathcal{I}_{m} =\{1,\ldots,m\}$ be an index set containing $m$ panels affected by the variance change, and
$\mathcal{I}_m^c=\{1,\ldots,N\}-\mathcal{I}_m=\{m+1,\ldots,N\}$ be the difference set.
Let $\mu_i=0$ and the time-varying scale parameters $(\sigma_{i}+\delta_iI(t>t^*))$ be specified as
\begin{equation*}
(\sigma_{i}+\delta_iI(t>t^*)) =
\begin{cases}
1, & t \le t^*, i\in \{1,\ldots,N\},\\
1 + \delta, & t > t^*, \; i \in \mathcal{I}_m, \\
1, & t > t^*, \; i \in \mathcal{I}_m^c.
\end{cases}
\end{equation*}

This construction corresponds to a \emph{sparse alternative}, where only a small fraction of panels experience a structural change.
In the simulations, we fix $N = 50, \quad T = 100, \quad t^* = 50, \quad m = 5$, so that only $10\%$ of the panels exhibit a variance change. The magnitude of the variance shift is set to $\delta = 0.8$, which represents a moderate but non-negligible deviation from the null hypothesis. All results are based on $1000$ Monte Carlo replications. For both test statistics, by the independent, the long-run variance component $s_i^{2*} = \sigma_i^4 \lim\limits_{T\to\infty}\frac{1}{T}\operatorname{Var}(\sum_{t=1}^Te_{i,t}^2)=\sigma_i^4\operatorname{Var}(e_{i,1}^2)$ is estimated using a simple variance estimator
$\hat{s}_{i,T}^{2*} = \frac{1}{T-1}\sum\nolimits_{t=1}^{T} (\hat{e}_{i,t}^2-\overline{ \hat{e}_i^2})^2$, where $\hat{e}_{i,t}=X_{i,t}-\overline{X}_i$, $\overline{X}_i=\frac{1}{T}\sum\nolimits_{t=1}^T X_{it}$ and $\overline{\hat{e}_i^{2}}=\frac{1}{T}\sum\nolimits_{t=1}^T \hat{e}_{i,t}^2$.

This choice is sufficient for illustrating the relative signal preservation properties of $T_V$ and $T_U$, and aligns with the asymptotic arguments in Section 3.

For each simulated dataset, we compute:
\begin{itemize}
\item $T_V$, the maximum absolute value of the panel-wise normalized CUSUM statistic proposed by \cite{li2015};
\item $T_U$, the proposed statistic that aggregates squared residuals across panels prior to normalization.
\end{itemize}
Both statistics are evaluated over all possible change-point locations $t=1,\ldots, T-1$.

\begin{figure}[htbp]
    \centering
    \includegraphics[width=0.95\textwidth]{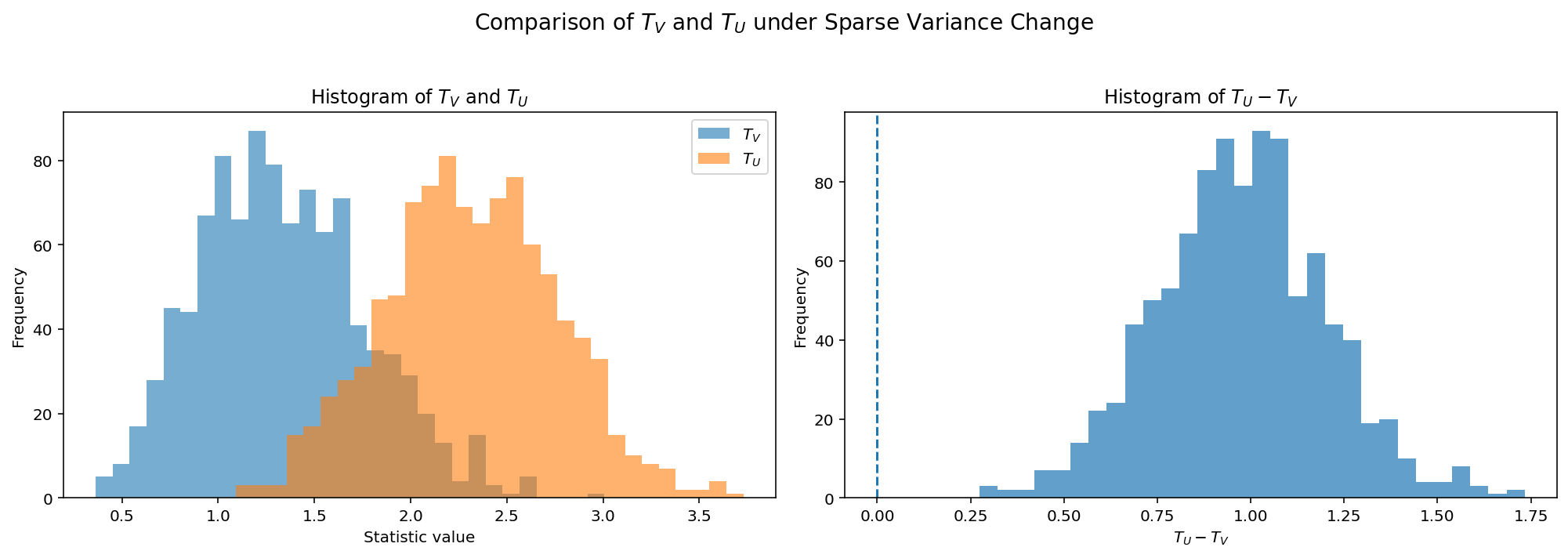}
    \caption{
    Comparison of the panel variance change statistics $T_V$ and $T_U$ under a sparse variance change model.
    The left panel displays the empirical distributions of $T_V$ and $T_U$ based on 1000 Monte Carlo replications.
    The right panel shows the distribution of the difference $T_U - T_V$.
    }
    \label{fig:TV_TU_sparse}
\end{figure}

Figure \ref{fig:TV_TU_sparse} compares the finite-sample behavior of the classical statistic $T_V$ and the proposed statistic $T_U$ under a \emph{sparse variance change} setting.  The results are based on $1000$ Monte Carlo replications.
In the left panel, the empirical distribution of $T_U$ is visibly shifted to the right relative to that of $T_V$, indicating that $T_U$ tends to take larger values when only a small subset of panels experiences a variance change. Quantitatively, the empirical signal-to-noise ratios are $\mathrm{SNR}(T_V) = 3.113$ and $\mathrm{SNR}(T_U) = 5.180$, showing that $T_U$ provides a stronger signal relative to noise compared to $T_V$.

The right panel plots the distribution of the difference $T_U-T_V$. The mass of the distribution lying entirely to the right of zero is consistent with the Monte Carlo estimate $\mathbb{P}(T_U > T_V) = 1.000$, providing direct evidence that $T_U$ dominates $T_V$ in detecting sparse variance changes.

Overall, the figure demonstrates that $T_U$, by effectively aggregating variance change information without dilution across unaffected panels, achieves higher SNR and greater sensitivity than $T_V$, making it particularly well-suited for detecting variance changes when only a few panels ($m \ll N$) are affected.

\section{Model Assumptions and Theoretical Results}

First, we state some assumptions in the panel data model \eqref{a1} as follows:
\begin{itemize}
\item[(A1)] Let $N\geq 1$. For each $i\in\{1,\ldots,N\}$, $\{e_{i,t},t\geq 1\}$ is a stationary sequence of $\alpha$-mixing with the mixing coefficient denoted by $\alpha_i(t)$. Let the sequences $\{e_{i,t},t\geq 1\},\ldots, \{e_{N,t},t\geq 1\}$ be independent of each other. For all $i\geq 1$ and $t\geq 1$, there are $Ee_{i,t}=0$, $\operatorname{Var}(e_{i,t})=1$ and $\sup\limits_{i\geq 1,t\geq 1}E|e_{i,t}|^{4+2\lambda}<\infty$ with some $\lambda>0$. Let $\sup\limits_{i\geq 1}\alpha_i(t)=O(t^{-\beta})$, where $\beta>\omega\frac{2+\lambda}{\lambda}$ and $\omega>1$.

 \item[(A2)]  For each $i\in \{1,2,\ldots,N\}$, let $\sigma_i^2$ be defined by \eqref{a1} satisfying
\begin{equation}
0<\inf\limits_{i\geq 1}\sigma_i^2\leq \sup\limits_{i\geq 1}\sigma_i^2<\infty.\label{b0}
\end{equation}

\item[(A3)] For each $i\in \{1,2,\ldots,N\}$, let
\begin{eqnarray}
\lim\limits_{T\rightarrow\infty}\frac{1}{T}\Big(\sum\limits_{t=1}^T\operatorname{Var}(e_{i,t}^2)+2\sum_{1\leq t_1<t_2\leq T}\operatorname{Cov}(
e_{i,t_1}^2,e_{i,t_2}^2)\Big)=s_{i}^2>0\label{b1}
\end{eqnarray}
and
\begin{equation}
0<\inf\limits_{i\geq 1}s_i^2\leq \sup\limits_{i\geq 1}s_i^2<\infty.\label{b1-1}
\end{equation}

\item[(A4)] Let
\begin{equation}
\frac{N}{T}\to 0, \label{b1-2}
\end{equation}
as $N,T\to \infty$.

\item[(A5)] For each $i\in \{1,\ldots,N\}$, let $\{e_{i,t},t\geq 1\}$ be a second-order stationarity sequence of $\alpha$-mixing with the mixing coefficient denoted by $\alpha_i(t)$. For each $i$, suppose $Ee_{i,1}=0$, $\operatorname{Var}(e_{i,1})=1$ and $Ee_{i,1}e_{i,1+j}^2=Ee_{i,1}^2e_{i,1+j}=0$ for all $j\geq 0$.
    Let the sequences $\{e_{i,t},t\geq 1\},\ldots, \{e_{N,t},t\geq 1\}$ be independent of each other. Assume that $\sup_{i\geq 1}E|e_{i,1}|^{8+4\lambda}<\infty$ for some $\lambda>0$. Let $\sup\limits_{i\geq 1}\alpha_i(t)=O(t^{-\beta})$, where $\beta>\omega\frac{2+\lambda}{\lambda}$ and $\omega>1$.

\item[(A6)]
For each $i\in \{1,2,\ldots,N\}$, let
\begin{equation}
s_{i}^2:=\gamma_{i}(0)+2\sum\limits_{h=1}^\infty\gamma_{i}(h)>0, \label{b3}
\end{equation}
where $\gamma_{i} (h)=Cov(e_{i,1}^2, e_{i,1+h}^2)$ for $h=0,1,2,\cdots$.

\item[(A7)]
Let $\{h_T, T\geq 1\}$ be a sequence of positive integers satisfying
\begin{equation}
h_T\rightarrow \infty ~~ \textrm{as}~~  T\rightarrow\infty  ~~\text{and} ~~ h_T=O(T^\eta)~~  \text{for ~some} ~\eta\in (0,1/2). \label{b4}
\end{equation}

\item[(A8)] Let $N\to \infty$, $T\to\infty$ and
\begin{equation}
N^{1/2}T^{-1/2+\eta}=o(1)~~\text{and}~~N^{1/2}T^{-(\omega-1)\eta}=o(1),\label{b5}
\end{equation}
where $\eta\in (0,1/2)$ is defined by \eqref{b4} and $\omega>1$ is defined by the condition (A5).
\end{itemize}

\textbf{Remark 3.1} \cite{li2015} considered the error $\{e_{i,t}\}$ to be a linear process:
$e_{i,t}=\sum_{j=1}^\infty c_{i,j}\varepsilon_{i,t-j}$, where $\varepsilon_{i,t-j}$ are $i.i.d.$ random variables and
$c_{i,j}$ are constant coefficients. Under some conditions of moments and coefficients, \cite{li2015} studies the variance change-point with the panel data model. In our paper, we consider the error to be $\alpha$-mixing error. For example, condition $(A1)$ is for the $\alpha$-mixing errors, which contains the conditions of moments and mixing coefficients. It will be used to study the asymptotic distribution for the CUSUM statistic \eqref{a4} based on $\hat{e}_{i,t}^2=(X_{i,t}-\overline{X}_i)^2$. In addition, the condition that $\{e_{i,t},t\geq 1\},\ldots, \{e_{N,t},t\geq 1\}$ are independent of each other can be relaxed to $m$-dependence, since the sequence of $m$-dependence can be decomposed into $m$ independent subsequences. Condition $(A2)$ requires the boundedness of error variances. The \eqref{b1} in $(A3)$ is used to study the limit variances of $\frac{1}{\sqrt{T}}\sum_{t=1}^{T}(e_{i,t}^2-1)$, $1\leq i\leq N$, and \eqref{b1-1} is the boundedness of these limit variances. For details, see the Lemma A.5 of Appendix A. Condition $(A4)$ means that the number $N$ of panels is smaller than the length of the observed time series $T$ (see \cite{horvan2012,li2015}). In order to estimate $s_{i}^2$ in $(A3)$, we need a second-order stationarity and high moment conditions in the $(A5)$, which are different from the $(A1)$.
We also need the conditions $Ee_{i,1}e_{i,1+j}^2=0$ and $Ee_{i,1}^2e_{i,1+j}=0$ to calculate $\operatorname{Var}\Big(\frac{1}{T}\sum_{t=1}^{T-h}e_{i,t}^2e_{i,t+h}\Big)$,
$\operatorname{Var}\Big(\frac{1}{T}\sum_{t=1}^{T}e_{i,t}^3\Big)$ and $\operatorname{Var}(\frac{1}{T}\sum_{t=1}^{T-h}e_{i,t}e_{i,t+h}^2)$ (see the proof of Lemma 3.1 for
estimating $s_{i}^{2*}=\sigma_i^4s_i^2$). Denote $f(x,y;k)$ as a joint probability density function of random variables $e_t$ and $e_{t+k}$ for all $t\geq 1$ and $k\geq 1$. Let $f(x,y;k)$ be symmetrical, i.e., $f(x,y;k)=f(-x,-y;k)$ for all $x,y\in R$. It is easy to check that
$E[e_te_{t+k}^2]=-Ee_te_{t+k}^2$ and $E[e_t^2e_{t+k}]=-Ee_t^2e_{t+k}$,
which implies $Ee_te_{t+k}^2=0$ and $Ee_t^2e_{t+k}=0$ for all $t\geq 1$ and $k\geq 1$. It also gets $Ee_1^3=0$.
For more details, see Remark 1 of \cite{gao2023}. The \eqref{b3} in $(A6)$ is a long-run variance of $s_{i}^2$. We use the sample autocovariance functions with condition $(A7)$ to estimate them (see \cite{lee2001,horvan2012,li2015}). For example, in \eqref{c1}, we can rewrite $\sigma_i^4s_i^2$ by $s_i^{2*}$ and give the estimator $\hat{s}_{i,T}^{2*}$ in \eqref{c4-2}. In the Lemma 3.1, we show that $\hat{s}_{i,T}^{2*}-s_{i}^{2*}=O(T^{-(\omega-1)\eta})+O_P(T^{-1/2+\eta})$, where $\omega>1$ and $\eta\in (0,1/2)$. This requires that the number $N$ of panels does not increase too large as $T$ increasing, i.e. $N^{1/2}T^{-1/2+\eta}=o(1)$ and $N^{1/2}T^{-(\omega-1)\eta}=o(1)$ in the condition $(A.8)$. If $\omega=2$ and $\eta=1/4$, then $N$ can take $T^{\kappa}$ with $\kappa\in (0,1/2)$. If one can improve the convergence rate for $\hat{s}_{i,T}^{2*}-s_{i}^{2*}$, the number $N$ can relax to be large. For more estimating of long-run variances, one can refer to \cite{lee2001}, \cite{horvan2012}, etc.

Next, we study the limit distribution for the CUSUM statistic $U_{N,T}(k)$ under the null hypothesis.

\textbf{Theorem 3.1}. Consider the change-point of variance in the panel data model \eqref{a1}. Let assumptions $(A1)$-$(A4)$ hold. Then, under $H_0$ defined by \eqref{a2}, we obtain
\begin{equation}
\frac{1}{\sqrt{T\sum_{i=1}^N\sigma_i^4s_{i}^2}}
\max\limits_{1\leq k\leq T-1}|U_{N,T}(k)|\stackrel{d}\rightarrow\sup\limits_{0\leq x\leq 1}|B^0(x)|\label{c1}
\end{equation}
where $U_{N,T}(k)$ is defined by \eqref{a4}, $\sigma_i^2$ and $s_{i}^2$ are
defined by \eqref{b0} and \eqref{b1}, respectively.

In the following, we discuss the estimator of $\sigma_i^4s_{i}^2$, $1\leq i\leq N$.
Denote
\begin{equation}
\sigma_i^4\operatorname{Cov}(e_{i,1}^2, e_{i,1+h}^2)=\sigma_i^4\gamma_i(h):=\gamma^{*}_i(h),h=0,1,2,\ldots,\label{c1-1}
\end{equation}
where $\gamma_i(h)=Cov(e_{i,1}^2, e_{i,1+h}^2)$ is defined by \eqref{b3}. Then $\sigma_i^4s_i^2$ can be rewritten by $s_i^{2*}$ as
\begin{equation}
s_i^{2*}:=\sigma_i^4s_i^2=\sigma_i^4\gamma_{i}(0)+2\sum\limits_{h=1}^\infty
\sigma_i^4\gamma_{i}(h):=\gamma^*_{i}(0)+2\sum\limits_{h=1}^\infty
\gamma^*_{i}(h),\label{c1-2}
\end{equation}
where $s_i^2$, $\gamma_i(h)$ and $\gamma^{*}_i(h)$ are respectively defined by \eqref{b3} and \eqref{c1-1}. So in order to estimate $s_i^{2*}$, we discuss the estimators of the sample autocovariance functions for $\gamma^{*}_i(h)$.
For example, $\gamma^*_{i}(h)$ can be estimated by
\begin{equation}
\hat{\gamma}^*_{i}(h)=\frac{1}{T}\sum\limits_{t=1}^{T-h} (\hat{e}_{i,t}^2-\overline{ \hat{e}_i^2})(\hat{e}_{i,t+h}^{2}-\overline{\hat{e}_i^2}),  ~  0\leq h<T, \label{c3}
\end{equation}
where $\hat{e}_{i,t}=X_{i,t}-\overline{X}_i$, $\overline{X}_i=\frac{1}{T}\sum\nolimits_{t=1}^T X_{it}$ and $\overline{\hat{e}_i^{2}}=\frac{1}{T}\sum\nolimits_{t=1}^T \hat{e}_{i,t}^2$. Thus, the estimator of $s_{i}^{2*}$ is suggested by
\begin{eqnarray}
\hat{s}_{i,T}^{2*}:=\hat{\gamma}^*_{i}(0)+2\sum\limits_{h=1}^{h_T}\hat{\gamma}^*_{i}(h), \label{c4-2}
\end{eqnarray}
where $\hat{\gamma}^*_{i}(h)$ and $h_T$ are defined by \eqref{c3} and \eqref{b4}, respectively.

Then, the consistency of convergence rate between $\hat{s}_{i,T}^{2*}$ and $s_i^{2*}$ is presented in Lemma 3.1.

\textbf{Lemma 3.1}. Consider the change-point of variance in the panel data model \eqref{a1}. Let the assumptions $(A5)$-$(A7)$ hold. Under $H_0$ defined by \eqref{a2}, for each $i\in \{1,2,\ldots,N\}$, it has
\begin{eqnarray}
\hat{s}_{i,T}^{2*}-s_{i}^{2*}=O(T^{-(\omega-1)\eta})+O_P(T^{-1/2+\eta}),\label{c5-2}
\end{eqnarray}
where $\omega>1$ and $\eta\in (0,1/2)$ are respectively defined by (A5) and \eqref{b4}, $s_{i}^{2*}$, $\hat{s}_{i,T}^{2*}$ and are defined by \eqref{c1-2} and \eqref{c4-2}, respectively.

Combining Theorem 3.1 with Lemma 3.1, one can get the following Theorem 3.2 immediately.

\textbf{Theorem 3.2}. Consider the change-point of variance in the panel data model \eqref{a1}. Let assumptions $(A5)$-$(A8)$ hold. Then, under $H_0$ defined by \eqref{a2}, we have
\begin{equation}
T_{U}=\frac{1}{\sqrt{T\sum_{i=1}^N\hat{s}_{i,T}^{2*}}}
\max\limits_{1\leq k\leq T-1}|U_{N,T}(k)|\stackrel{d}\rightarrow\sup\limits_{0\leq x\leq 1}|B^0(x)|\label{c6}
\end{equation}
where $U_{N,T}(k)$ is defined by \eqref{a4}, and $\hat{s}_{i,T}^{2*}$ is defined by \eqref{c4-2}.

Last, we consider the limit behavior for the CUSUM statistic $U_{N, T}(k)$ under the alternative hypothesis.

\textbf{Theorem 3.3}. Let assumptions $(A1)$-$(A4)$ hold. Consider the change-point of variance in the panel data model \eqref{a1} satisfying
\begin{eqnarray}
0<\liminf\limits_{T\to\infty}\frac{t^*}{T}\leq \limsup\limits_{T\to\infty}\frac{t^*}{T}<1,\label{c7}\\
\min_{1\leq i\leq N}(\sigma_i+\delta_i)>0,\label{c7*}\\
\frac{N}{T}\frac{\max_{1\leq i\leq N}|\delta_i|}{ \left||\sum_{i=1}^N\sigma_i\delta_i|-\sum_{i=1}^N\delta_i^2\right|}\to 0,~~\min(N,T)\to \infty,\label{c7**}\\
\frac{1}{\sqrt{T}}\frac{|\sum_{i=1}^N\sigma_i\delta_i|+\sum_{i=1}^N\delta_i^2}{\left||\sum_{i=1}^N\sigma_i\delta_i|-\sum_{i=1}^N\delta_i^2\right|}\to 0,~~\min(N,T)\to \infty,\label{c8}\\
\frac{\sqrt{T}}{\sqrt{N}}\left|\Big|\sum_{i=1}^N\sigma_i\delta_i\Big|-\sum_{i=1}^N\delta_i^2\right|\to\infty,~\min(N,T)\to \infty.\label{c9}
\end{eqnarray}
Then
\begin{equation}
\frac{1}{\sqrt{T\sum_{i=1}^N\sigma_i^4s_{i}^2}}
\max\limits_{1\leq k\leq T-1}|U_{N,T}(k)|\stackrel{P}\rightarrow\infty. \label{c10}
\end{equation}

\textbf{Remark 3.2}. \cite{li2015} considered the change-point of variance in panel data model \eqref{panel_li} with linear process of errors, where $\delta_i=0$ for all $1\leq i\leq N$. They gave the CUSUM statistic $V_{N, T}(k)$ defined by \eqref{V_stat_li}. For the estimator $\hat{s}_{i,T}^{2*}$ in \eqref{V_stat_li}, \cite{li2015} used the method of \cite{lee2001} to show that there existed a positive constant $s_i^{2*}$ satisfying $\hat{s}_{i,T}^{2*}\stackrel{P}\rightarrow s_i^{2*}$, as $T\to\infty$.
Under the null hypothesis of no variance change, for $1\leq i\leq N$, they also gave that
$$\max\limits_{1\leq k\leq T-1}\frac{1}{\sqrt{\hat{s}_{i,T}^{2*}}}\frac{1}{\sqrt{T}}\Big|\sum_{t=1}^ke_{i,t}^2-\frac{k}{T}\sum_{t=1}^Te_{i,t}^2\Big|\stackrel{d}\rightarrow\sup\limits_{0\leq x\leq 1}|B^0(x)|,~~~~1\leq i\leq N$$ and
\begin{equation}
T_V=\max\limits_{1\leq k\leq T-1}\frac{1}{\sqrt{NT}}\Big|\sum_{i=1}^N\frac{1}{\sqrt{\hat{s}_{i,T}^{2*}}}\Big(\sum_{t=1}^k\hat{e}_{i,t}^2-\frac{k}{T}\sum_{t=1}^T\hat{e}_{i,t}^2\Big)\Big|
\stackrel{d}\rightarrow\sup\limits_{0\leq x\leq 1}|B^0(x)|,\label{cr3}
\end{equation}
where $e_{i,t}=X_{i,t}-\mu_i$ and $\hat{e}_{i,t}=X_{i,t}-\overline{X}_i$. It extended the results of \cite{horvan2012}. But they did not give the
consistency rate between $\hat{s}_{i,T}^{2*}$ and $s_{i,T}^{2*}$. In our Lemma 3.1, we give the consistency rate $\hat{s}_{i,T}^{2*}-s_{i}^{2*}=O(T^{-(\omega-1)\eta})+O_P(T^{-1/2+\eta})$. The difference between $T_U$ test and $T_V$ test is that our test $T_U$ contains a formula
$\frac{1}{\sqrt{\sum_{i=1}^N\hat{s}_{i,T}^{2*}}}$, while \cite{li2015}'s test $T_V$ contains a formula
$\frac{1}{\sqrt{N}}\frac{1}{\sqrt{\hat{s}_{i,T}^{2*}}}$. In fact, by the proof of Theorem 3.1, $\frac{1}{\sqrt{T\sum_{i=1}^N\hat{s}_{i,T}^{2*}}}U_{N,T}(k)$ is absolutely integrable.
If $\frac{1}{\sqrt{N}}\frac{1}{\sqrt{\hat{s}_{i,T}^{2*}}}$ in \eqref{V_stat_li} is replaced by $\frac{1}{\sqrt{\sum_{i=1}^N\hat{s}_{i,T}^{2*}}}$, then $V_{N,T}(k)$ can be be rewritten by $V_{N,T}^*(k)$ as
\begin{eqnarray}
V_{N,T}^*(k)&=&\sum_{i=1}^N\frac{1}{\sqrt{\sum_{i=1}^N\hat{s}_{i,T}^{2*}}}\frac{1}{\sqrt{T}}\Big(\sum_{t=1}^k\hat{e}_{i,t}^2-\frac{k}{T}\sum_{t=1}^T\hat{e}_{i,t}^2\Big)\nonumber\\
&=&\frac{1}{\sqrt{T\sum_{i=1}^N\hat{s}_{i,T}^{2*}}}\frac{k(T-k)}{T}\Big(\frac{1}{k}\sum_{t=1}^k\sum_{i=1}^N\hat{e}_{i,t}^2-\frac{1}{T-k}\sum_{t=1+k}^{T}\sum_{i=1}^N\hat{e}_{i,t}^2\Big)
\nonumber\\
&=&\frac{1}{\sqrt{T\sum_{i=1}^N\hat{s}_{i,T}^{2*}}}U_{N,T}(k).\nonumber
\end{eqnarray}
So it will have our test, i.e.
\begin{equation}
T_{V}^*=\max\limits_{1\leq k\leq T-1}|V_{N,T}^*(k)|=\frac{1}{\sqrt{T\sum_{i=1}^N\hat{s}_{i,T}^{2*}}}
\max\limits_{1\leq k\leq T-1}|U_{N,T}(k)|=T_{U}.\nonumber
\end{equation}

Under the null hypothesis, the tests $T_U$ and $T_V$ have the same limit distribution.
However, under the alternative hypothesis, the limit behavior is not the same. In fact, for each $i\in \{1,2,\ldots,N\}$,
the estimator $\hat{s}_{i,T}^{2*}$ in $T_V$ test increases by a factor $N$, while the estimator $\hat{s}_{i,T}^{2*}$ in our $T_U$ test has a harmonized formula $\sum_{i=1}^N\hat{s}_{i,T}^{2*}$. If all the estimators $\hat{s}_{i,T}^{2*}$ have the same order, then there may be $N\hat{s}_{i,T}^{2*}\approx\sum_{i=1}^N\hat{s}_{i,T}^{2*}$, i.e. the tests $T_U$ and $T_V$ may have the same limit behavior. But in more situations, $N\hat{s}_{i,T}^{2*}$ and $\sum_{i=1}^N\hat{s}_{i,T}^{2*}$ are not equal in the asymptotic sense. In other words, if there exist some changes of variance in the panel data model, then the consistency of $\hat{s}_{i, T}^{2*}$ may be changed, and
significant differences may exist between $N\hat{s}_{i, T}^{2*}$ and $\sum_{i=1}^N\hat{s}_{i, T}^{2*}$.
For example, in the panel data model \eqref{a1}, for each $i\in \{1,2,\ldots, N\}$, if $t^*=T$ (i.e., there is no change of variance), then by \eqref{c4-2} and Lemma 3.1, we have
$$\hat{s}_{i,T}^{2*}=\hat{\gamma}^*_{i}(0)+2\sum\limits_{h=1}^{h_T}\hat{\gamma}^*_{i}(h)\stackrel{P}\rightarrow \sigma_i^4s_i^2:=s^{2*}_i,~~T\to\infty,$$
where
$\hat{\gamma}^*_{i}(h)=\frac{1}{T}\sum\nolimits_{t=1}^{T-h} (\hat{e}_{i,t}^2-\overline{ \hat{e}_i^2})(\hat{e}_{i,t+h}^{2}-\overline{\hat{e}_i^2})$, $\hat{e}_{i,t}^2$ and $\overline{ \hat{e}_i^2}$ are defined by \eqref{c3},
and $s_i^2$ is defined by \eqref{b3}. Similarly, if $t^*=0$ in \eqref{a1}, for each $i$, there is
$$\hat{s}_{i,T}^{2*}\stackrel{P}\rightarrow (\sigma_i+\delta_i)^4s_i^2:=s^{2**}_i,~~T\to\infty.$$
Let $t^*\in \{1,2,\ldots,T-1\}$ be a change-point of variance in the panel data model \eqref{a1} satisfying $t^*/T\to \omega\in (0,1)$ as $T\to\infty$. Then, by the weighted average method, one will have
$$\hat{s}_{i,T}^{2*}=\hat{\gamma}^*_{i}(0)+2\sum\limits_{h=1}^{h_T}\hat{\gamma}^*_{i}(h)
\stackrel{P}\rightarrow \bar{s}^{2*}_i,~~T\to\infty,$$
where
\begin{eqnarray}
\hat{\gamma}^*_{i}(h)&=&\frac{1}{T}\sum\limits_{t=1}^{T-h} (\hat{e}_{i,t}^2-\overline{ \hat{e}_i^2})(\hat{e}_{i,t+h}^{2}-\overline{\hat{e}_i^2})\nonumber\\
&=&\frac{t^*}{T}\frac{1}{t^*}\sum\limits_{t=1}^{t^*} (\hat{e}_{i,t}^2-\overline{ \hat{e}_i^2})(\hat{e}_{i,t+h}^{2}-\overline{\hat{e}_i^2})
+\frac{T-h-t^*}{T}\frac{1}{T-h-t^*}\sum\limits_{t=t^*+1}^{T-h} (\hat{e}_{i,t}^2-\overline{ \hat{e}_i^2})(\hat{e}_{i,t+h}^{2}-\overline{\hat{e}_i^2}),\nonumber
\end{eqnarray}
$\hat{e}_{i,t}^2$ and $\overline{ \hat{e}_i^2}$ are defined by \eqref{c3},
and $\bar{s}^{2*}_i$ is a positive constant between $s^{2*}_i$ and $s^{2**}_i$, $1\leq i\leq N$.
So, for each $i$, $\hat{s}_{i,T}^{2*}$ in the test $T_V$ increases to $N\hat{s}_{i,T}^{2*}$, while $\hat{s}_{i,T}^{2*}$ in our test $T_U$ increases to $\sum_{i=1}^N\hat{s}_{i,T}^{2*}$. We find that the performances of our test $T_U$ are better than the ones of test $T_V$,
particularly in sparse changes of variance. For example, let's consider a simple case. In the model \eqref{a1}, let $\sigma_1=\ldots=\sigma_N=\sigma_0>0$ and $\{e_{i,t},1\leq i\leq N,1\leq t\leq T\}$ $i.i.d.$ random variables with $Ee_{1,1}=0$ and $Ee_{1,1}^2=1$. Let $m$ be a positive integer much smaller than $N$. We consider the sparse changes in variance and assume that there are only $m$ signal changes of variance such as $\delta_1=\ldots=\delta_m=\delta_0>0$ and $\delta_{m+1}=\ldots=\delta_N=0$. Then, for $1\leq i\leq m$, $\hat{s}_{i,T}^{2*}\stackrel{P}\rightarrow \bar{s}^4$,
where $\sigma_0^4<\bar{s}^4<(\sigma_0+\delta_0)^4$. Meanwhile, for $m+1\leq i\leq N$,
$\hat{s}_{i,T}^{2*}\stackrel{P}\rightarrow \sigma_0^4$. Thus, in probability, we have
$$\sum_{i=1}^N\hat{s}_{i,T}^{2*}=\sum_{i=1}^m\hat{s}_{i,T}^{2*}+\sum_{i=m+1}^N\hat{s}_{i,T}^{2*}\approx m\bar{s}^4+(N-m)\sigma_0^4,$$
and
$$N\hat{s}_{i,T}^{2*}\approx N\bar{s}^4 >m\bar{s}^4+(N-m)\sigma_0^4, ~~1\leq i\leq m.$$
Consequently, for $1\leq i\leq m$, the signal change $\delta_i>0$ of variance, $\frac{1}{\sqrt{N}}\frac{1}{\sqrt{\hat{s}_{i,T}^{2*}}}$ in $T_V$ is smaller in probability than $\frac{1}{\sqrt{\sum_{i=1}^N\hat{s}_{i,T}^{2*}}}$ in $T_U$, which reduces the power of test $T_V$.

That's why in this paper, we also consider the change-point of variance in the panel data model and give a more powerful test $T_U$ defined by \eqref{TU_stat}. See the comparison in Figure \ref{fig:TV_TU_sparse}.
In the simulation of Section 4, we also consider the spares changes of variance with $\delta_i>0$ and $\delta_i<0$. For example, we randomly select 10 values for $\delta_i$, with 5 set to 1.5, 5 set to -0.5, and all others set to zero;
The power of our test $T_U$ is close to 1, while the power of the test $T_V$ is much smaller than 0.1.
We also randomly select 10 $\delta_i$ values as -0.5, and set all others to zero. It shows that the power of our test $T_U$ is also better than the one of test $T_V$.

\textbf{Remark 3.3}.
In the panel data model \eqref{panel_li} with linear process of errors and $\delta_i=0$ for all $1\leq i\leq N$, \cite{li2015} considered the variance of panel $i$ changes from $\sigma_i^2$ to $\sigma_i^2+\delta_i$ with $\delta_i>0$. When $\frac{T^{1/2}}{N^{1/2}}\sum_{i=1}^N\delta_i\to \infty$ as $\min(N,T)\to \infty$, they showed that $\max\limits_{1\leq k\leq T-1}|V_{N,T}(k)|\stackrel{P}\rightarrow\infty$, which extended Theorem 3 of \cite{horvan2012} for the mean change-point to the variance change-point.
We consider the change-point of variance in the panel data model \eqref{a1}, which is different from the one of \cite{li2015}.
Next, we give two examples to illustrate the changes of variance for the conditions of \eqref{c7*}, \eqref{c7**}, \eqref{c8} and \eqref{c9}. For simplicity, let $\sigma_1=\ldots=\sigma_N=\sigma_0>0$ and $N/T=o(1)$. Let $\delta_i=i^{-1/2}$, $1\leq i\leq N$.
It is easy to see that $c_1N^{1/2}\leq \sum_{i=1}^N\delta_i\leq c_2N^{1/2}$ and $c_3\log N\leq \sum_{i=1}^N\delta_i^2\leq c_4\log N$,
$\Big|\sum_{i=1}^N\delta_i-\sum_{i=1}^N\delta_i^2\big|\geq c_5N^{1/2} $, $\max\limits_{1\leq i\leq N}|\delta_i|=O(1)$,
where $c_1,\ldots,c_5$ are positive constants. So
\begin{eqnarray}
\frac{N}{T}\frac{\max_{1\leq i\leq N}|\delta_i|}{ \Big||\sum_{i=1}^N\sigma_i\delta_i|-\sum_{i=1}^N\delta_i^2\Big|}\leq c_6\frac{N}{TN^{1/2}}=O(\frac{N^{1/2}}{T})=o(1),\nonumber\\
\frac{1}{\sqrt{T}}\frac{|\sum_{i=1}^N\sigma_i\delta_i|+\sum_{i=1}^N\delta_i^2}{\Big||\sum_{i=1}^N\sigma_i\delta_i|-\sum_{i=1}^N\delta_i^2\Big|}
\leq c_7\frac{1}{\sqrt{T}}=o(1),\nonumber\\
\frac{\sqrt{T}}{\sqrt{N}}\left|\Big|\sum_{i=1}^N\sigma_i\delta_i\Big|-\sum_{i=1}^N\delta_i^2\right|
\geq c_8\frac{\sqrt{T}}{\sqrt{N}}N^{1/2}=c_8\sqrt{T}\to\infty.\nonumber
\end{eqnarray}
Similarly, let $\delta_i=i^{1/2}$, $1\leq i\leq N$.
It has $$\max\limits_{1\leq i\leq N}|\delta_i|=O(N^{1/2}),~~c_9N^{3/2}\leq \sum_{i=1}^N\delta_i\leq c_{10}N^{3/2},~~c_{11}N^2\leq \sum_{i=1}^N\delta_i^2\leq c_{12}N^2,$$
$$\Big|\sum_{i=1}^N\delta_i-\sum_{i=1}^N\delta_i^2\Big|\geq c_{13}N^{3/2}.$$ Thus
\begin{eqnarray}
\frac{N}{T}\frac{\max_{1\leq i\leq N}|\delta_i|}{ \Big||\sum_{i=1}^N\sigma_i\delta_i|-\sum_{i=1}^N\delta_i^2\Big|}\leq c_{14}\frac{N^{3/2}}{TN^{3/2}}=O(\frac{1}{T})=o(1),\nonumber\\
\frac{1}{\sqrt{T}}\frac{|\sum_{i=1}^N\sigma_i\delta_i|+\sum_{i=1}^N\delta_i^2}{\Big||\sum_{i=1}^N\sigma_i\delta_i|-\sum_{i=1}^N\delta_i^2\Big|}
\leq c_{15}\frac{N^2}{\sqrt{T}N^{3/2}}O(\sqrt{\frac{N}{T}})=o(1),\nonumber\\
\frac{\sqrt{T}}{\sqrt{N}}\left|\Big|\sum_{i=1}^N\sigma_i\delta_i\Big|-\sum_{i=1}^N\delta_i^2\right|
\geq c_{16}\frac{\sqrt{T}}{\sqrt{N}}N^{3/2}=c_{16}\sqrt{T}N\to\infty.\nonumber
\end{eqnarray}
Therefore, the conditions of \eqref{c7*}, \eqref{c7**}, \eqref{c8} and \eqref{c9} are satisfied. So we obtain $\frac{1}{\sqrt{T\sum_{i=1}^N\sigma_i^4s_{i}^2}}\max\limits_{1\leq k\leq T-1}|U_{N,T}(k)|\stackrel{P}\rightarrow\infty$ in the Theorem 3.3.

\setcounter{equation}{0}
\section{Simulations}

In this section, we do some simulations to show the performances of our test $T_U$ and other test $T_V$.
We consider a single change-point of variance in the panel data model \eqref{a1} such as
\begin{equation}
X_{i,t}=\mu_i+(\sigma_i+\delta_iI(t\geq t^*+1))e_{i,t}, 1\leq i\leq N,~~1\leq t\leq T.\label{e1}
\end{equation}

First, we consider the error term. Let $\boldsymbol{e}^i=(e_{i,1},e_{i,2},\ldots,e_{i,T})\stackrel{i.i.d.}\sim(\varepsilon_1,\varepsilon_2,\ldots,\varepsilon_T)$, $1\leq i\leq N$.
Assume that $(\varepsilon_1,\varepsilon_2,\ldots,\varepsilon_T)$ satisfies a general multivariate model
\begin{equation}
(\varepsilon_1,\varepsilon_2,\ldots,\varepsilon_T)=(\tilde{\varepsilon}_1,\tilde{\varepsilon}_2,\ldots,\tilde{\varepsilon}_T)\boldsymbol{\Gamma}^\prime\nonumber
\end{equation}
where the entries $\tilde{\varepsilon}_2,\tilde{\varepsilon}_2,\ldots,\tilde{\varepsilon}_T$ are $i.i.d.$ random variables with $E\tilde{\varepsilon}_1=0$, $\operatorname{Var}(\tilde{\varepsilon}_1)=1$, and $\boldsymbol{\Gamma}^\prime\boldsymbol{\Gamma}$ is a Toeplitz covariance matrix \citep{Gray1972}
satisfying
\begin{equation}
\boldsymbol{\Gamma}^\prime\boldsymbol{\Gamma}=\begin{bmatrix}
\rho_0 & \rho_1,&\ldots,&\rho_{T-1}\\
\rho_1 & \rho_0,&\ldots,&\rho_{T-2}\\
\ldots,&\ldots,&\ldots,&\ldots\\
\rho_{T-1} & \ldots,&\rho_1,&\rho_{0}
\end{bmatrix}.\label{e2}
\end{equation}
Here $\rho_i$ takes two cases such as $\rho_i=2^{-i}$ and $\rho_i=i^{-2}$ for $i=0,1,\ldots,T-1$.
When $\rho_i=2^{-i}$ or $\rho_i=i^{-2}$ in \eqref{e2}, we respectively take $h_T=\lfloor T^{1/4}\rfloor$ and
$h_T=\lfloor T^{1/3}\rfloor$ in the process of estimator $\hat{s}_{i,T}^{2*}$ by \eqref{c4-2}.
In the simulation, we consider $\tilde{\varepsilon}_1$ to be \textbf{Gaussian} and  \textbf{Gamma} random variables such as
$\tilde{\varepsilon}_1\sim N(0,1)$ and $\tilde{\varepsilon}_1\sim \frac{Gamma(4, 1)-4}{2}$.

Second, we consider the panel data model \eqref{e1} with a matrix form.
Let $A\odot B$ be a Hadamard product. For example, if $A,B\in R^2$, then
\begin{equation}
A\odot B=\begin{bmatrix}
a_{11} & a_{12}\\
a_{21} & a_{22}
\end{bmatrix}\odot\begin{bmatrix}
b_{11} & b_{12}\\
b_{21} & b_{22}
\end{bmatrix}=\begin{bmatrix}
a_{11}b_{11} & a_{12}b_{12}\\
a_{21}b_{21} & a_{22}b_{22}
\end{bmatrix}.\nonumber
\end{equation}
Then, the panel data model \eqref{e1} can be written by
\begin{equation}
\boldsymbol{X}_{N\times T}=\boldsymbol{\mu}_{N\times T}+\boldsymbol{\Delta}_{N\times T}\odot \boldsymbol{e}_{N\times T},\label{e3}
\end{equation}
where
\begin{equation}
\boldsymbol{\mu}_{N\times T}=\begin{bmatrix}
\mu_1 & \mu_1,&\ldots,&\mu_{1}\\
\mu_2 & \mu_2,&\ldots,&\mu_{2}\\
\ldots,&\ldots,&\ldots,&\ldots\\
\mu_N & \mu_N,&\ldots,&\mu_{N}
\end{bmatrix},~~~
\boldsymbol{e}_{N\times T}=\begin{bmatrix}
\boldsymbol{e}^1\\
\vdots\\
\boldsymbol{e}^N\\
\end{bmatrix},\nonumber
\end{equation}
\begin{equation}
\boldsymbol{\Delta}_{N\times T}=\begin{bmatrix}
\sigma_1, &  \ldots,&\sigma_1,&\sigma_1+\delta_1,&\ldots,&\sigma_1+\delta_1\\
\sigma_2, & \ldots,&\sigma_2,&\sigma_2+\delta_2,&\ldots,&\sigma_2+\delta_2\\
\ldots,&\ldots,&\ldots,&\ldots,&\ldots,&\ldots\\
\sigma_N, & \ldots,&\sigma_N,&\sigma_N+\delta_N,&\ldots,&\sigma_N+\delta_N
\end{bmatrix},\nonumber
\end{equation}
and $(\underbrace{\sigma_1, \ldots,\sigma_1}_{t^*},\underbrace{\sigma_1+\delta_1,\ldots,\sigma_1+\delta_1}_{T-t^*})$. In the simulation, we take the change-point location by $t^*=\lfloor cT\rfloor$ with $c=1/3$ and $c=1/2$, where $\lfloor x\rfloor$ denotes the largest integer not exceeding $x$.

Third, we consider the terms of $\mu$, $\delta$ and $\sigma$. Let $\mu_1,\ldots,\mu_N$ be $i.i.d.$ random variables with $\mu_1\sim U(0,1)$. Similarly, let $\sigma_1,\ldots,\sigma_N$ be $i.i.d.$ random variables with $\sigma_1\sim U(1,2)$. Next, we consider the changes of variance $\delta_i$. In the following, we take them as \textbf{non-sparse} and  \textbf{sparse} cases:
\begin{itemize}
\item[1)] Non-sparse cases: let $\delta_1,\ldots,\delta_N$ be $i.i.d.$ random variables with  \textbf{(a)} $\delta_1\sim U(-0.5,0.5)$; \textbf{(b)}
$\delta_1\sim U(-0.5,1)$.
\item[2)] Sparse cases: \textbf{(a)}  randomly select 10 values for $\delta_i$, with 5 set to 1.5, 5 set to -0.5, and all others set to zero; \textbf{(b)} randomly select 10 $\delta_i$ values as -0.5, and set all others to zero.
\end{itemize}

Fourth, we discuss the accuracy of a change-point estimator $\hat{t}_{N, T}$.
Let $\alpha$ be the significance level $C_\alpha$ be the threshold defined by Remark 3.1.
An accuracy of the estimator $\hat{t}_{N,T}$ for the variance change-point based on
test $T_{test}$ is also suggested by
$$Accuracy=\frac{\sharp\{T_{U}\geq C_\alpha~~\text{and}~~|\hat{t}_{N,T}-t^*|\leq m \}}{M},$$
where $m$ is a positive constant, $M$ is the number of simulations and $\sharp\{A\}$ denotes the length or size of set $A$. In the simulation, $\alpha$ and $C_\alpha$ are respectively taken by 0.05 and 1.358,
$m$ and $M$ are respectively taken by $0.05T$ and 1000. We also take the individual number $N=[100,200,300,400,1000,1500,2000]$ and length time $T=[500,1000,2000,4000]$. \cite{Choi2021} used a self-normalization
method to have a variance change test. So as a comparison, we denote our test by $T_U$, \cite{li2015}'s test by $T_V$ and \cite{Choi2021}'s test by $T_{LV}$, respectively.

Under the null hypothesis $H_0$ of no variance change in the panel data model \eqref{e3},
we give the empirical size of the tests $T_U$, $T_V$ and $T_{LV}$ based on the Gaussian and Gamma errors in the Table \ref{tab1}.
\begin{table}[htbp]
\begin{center}
\caption{\footnotesize Empirical size of tests $T_U$, $T_V$ and $T_{LV}$ with different errors and significance level $\alpha=0.05$}\label{tab1}
\vskip1mm
\scriptsize
\begin{tabular}{|c|cccccc|cccccc|}
\hline
$\rho$&Error& N& T& size $T_U$& size $T_V$&size $T_{LV}$&Error& N& T& size $T_U$& size $T_V$&size $T_{LV}$\\
\hline
$2^{-i}$&Gaussian&	100&	500	&0.036&	0.044&	0.054&Gamma&	100&	500	&0.045&	0.043 &0.056\\
$2^{-i}$&Gaussian&	200	&1000&	0.041&	0.043&	0.053&Gamma&	200&	1000&	0.049&	0.043&0.057\\
$2^{-i}$&Gaussian&	300	&2000&	0.061&	0.047&	0.047&Gamma&	300&	2000&	0.047&	0.034&0.056\\
$2^{-i}$&Gaussian&	400&	2000&	0.043&	0.051&	0.067&Gamma&	400&	2000&	0.038&	0.039&0.053\\
$2^{-i}$&Gaussian&	1000&	4000&	0.043&	0.046&	0.055&Gamma&	1000&	4000&	0.049&	0.044&0.068\\
$2^{-i}$&Gaussian&	1500&	4000&	0.054&	0.058&	0.048&Gamma&	1500&	4000&	0.053&	0.048&0.056\\
$2^{-i}$&Gaussian&	2000&	4000&	0.052&	0.055&	0.054&Gamma&	2000&	4000&	0.053&	0.053&0.059\\
\hline
$i^{-2}$	&Gaussian&	100	&500	&0.044&	0.034&	0.046&Gamma&	100&	500	&0.036&	0.041&0.047\\
$i^{-2}$&	Gaussian	&200	&1000&	0.055&	0.047&	0.054&Gamma&	200	&1000&	0.038&	0.043&0.054\\
$i^{-2}$	&Gaussian&	300&	2000&	0.043&	0.039&	0.055&Gamma	&300&	2000&	0.055&	0.060&0.047\\
$i^{-2}$& Gaussian&	400&	2000&	0.042&	0.044&	0.051&Gamma	&400&	2000	&0.049&	0.049&0.065\\
$i^{-2}$& Gaussian	&1000&	4000&	0.051&	0.053&	0.054&Gamma&	1000&	4000&	0.056&	0.045&0.050\\
$i^{-2}$	&Gaussian&	1500	&4000&	0.046&	0.049&	0.051&Gamma&	1500&	4000&	0.051&	0.045&0.051\\
$i^{-2}$	&Gaussian&	2000&	4000&	0.046&	0.044&	0.060&Gamma&	2000&	4000&	0.047&	0.048&0.049\\
\hline
\end{tabular}
\end{center}
\end{table}

By the Table \ref{tab1}, it can be seen that all of our test $T_U$, \cite{li2015}'s test $T_V$, and \cite{Choi2021}'s test $T_{LV}$ control the empirical size at the significance level $\alpha=0.05$.

Next, we consider the tests $T_U$, $T_V$ and $T_{LV}$ for the alternative hypothesis $H_1$ of variance change in the panel data model.
For the change point location $t^*=\lfloor T/2\rfloor$, the empirical power and accuracy of tests $T_U$, $T_V$ and $T_{LV}$ with Gaussian errors are presented in the Table \ref{tab2}.
\begin{table}[htbp]
\begin{center}
\caption{\footnotesize Empirical power and accuracy of tests $T_U$, $T_V$ and $T_{LV}$ with Gaussian error, significance level $\alpha=0.05$ and location $t^*=\lfloor T/2\rfloor$}\label{tab2}
\vskip1mm
\scriptsize
\renewcommand{\arraystretch}{0.6}
\begin{tabular}{|cccccccccc|}
\hline
$\rho$&$\delta$& N& T& power $T_U$& power $T_V$&power $T_{LV}$ &accuracy $T_U$& accuracy $T_V$&accuracy $T_{LV}$\\
\hline
$2^{-i}$ &	$U(-0.5,0.5)$&  100	&  500&  	0.494&	0.455&	0.522& 0.323&	0.316&0.320\\
$2^{-i}$ &	$U(-0.5,0.5)$&	200 &  1000&	0.708&	0.651&	0.698&0.577&	0.531&0.535\\
$2^{-i}$ &  $U(-0.5,0.5)$&	300	&  2000&    0.859&	0.788&	0.844&0.776&	0.703&0.728\\
$2^{-i}$ &  $U(-0.5,0.5)$&	400 &  2000&	0.904&	0.826&	0.903&0.836&	0.745&0.780\\
$2^{-i}$ &  $U(-0.5,0.5)$&	1000&  4000&	0.999&	0.967&	0.989&0.996&	0.953&0.979\\
$2^{-i}$ &	$U(-0.5,0.5)$&	1500&  4000&	0.999&	0.990&	0.999& 0.998&	0.980&0.998\\
$2^{-i}$ &  $U(-0.5,0.5)$&	2000&  4000&	1    &	0.995&	1&1    &	0.995&0.998\\
\hline
$2^{-i}$ &	$U(-0.5,1)$	&100   &500    &	1&	1&	0.999& 1&0.987&0.993\\
$2^{-i}$ &	$U(-0.5,1)$ &	200&	1000&	1&	1&  1&	1 &1 &1\\
$2^{-i}$ &  $U(-0.5,1)$ &	300	&2000	&   1&	1&	1&	1 &1 &1\\
$2^{-i}$ &  $U(-0.5,1$ &	400&	2000&	1&	1&	1&	1 &1 &1\\
$2^{-i}$ &  $U(-0.5,1)$&	1000&	4000&	1&	1&	1&	1 &1 &1\\
$2^{-i}$ &	$U(-0.5,1)$&	1500&	4000&	1&	1&	1&	1 &1 &1\\
$2^{-i}$ &  $U(-0.5,1)$&	2000&	4000&	1&	1&	1&	1 &1 &1\\
\hline
$2^{-i}$    &Sparse (a)&    100	   &500  &	\bf{1}  &	0.047&0.087&	\bf{0.915}&	0.014&0.031\\
$2^{-i}$    &Sparse (a)&	200    & 1000&	\bf{1}  &	0.066&0.096&	\bf{0.933}&	0.013&0.031\\
$2^{-i}$ 	&Sparse (a)&	300	   &2000	&\bf{1} &	0.074&0.112&	\bf{0.955}&	0.025&0.037\\
$2^{-i}$ 	&Sparse (a)&	400   &	2000&	\bf{1}&	0.063&	0.106&\bf{0.931}&	0.017&0.026\\
$2^{-i}$ 	&Sparse (a)&	1000  &	4000&	\bf{1}&	0.062&	0.093&\bf{0.906}&	0.022&0.027\\
$2^{-i}$    &Sparse (a)&	1500  &	4000&	\bf{0.991}&	0.043&	0.064&\bf{0.833}&	0.011&0.019\\
$2^{-i}$ 	&Sparse (a)&	2000  &	4000&	\bf{0.969}&	0.049&	0.071&\bf{0.758}&	0.015&0.016\\
\hline
$2^{-i}$    &Sparse (b)&    100	   &500  &	0.801&	0.747&0.908	&0.536&	0.422&0.631\\
$2^{-i}$    &Sparse (b)&	200    & 1000&	0.798&	0.618&0.911&0.538&	0.355&0.646\\
$2^{-i}$ 	&Sparse (b)&	300	   &2000&	0.877&	0.593&0.956&0.616&	0.337&0.731\\
$2^{-i}$ 	&Sparse (b)&	400   &	2000&	0.778&	0.453&0.907&0.512&	0.232&0.643\\
$2^{-i}$ 	&Sparse (b)&	1000  &	4000&	0.678&	0.301&0.828&0.431&	0.153&0.527\\
$2^{-i}$    &Sparse (b)&	1500  &	4000&	0.509&	0.223&0.690&0.269&	0.082&0.380\\
$2^{-i}$ 	&Sparse (b)&	2000  &	4000&	0.430&	0.163&0.573&0.215&	0.068&0.302\\
\hline
$i^{-2}$    &$U(-0.5,0.5)$	&100       &500 &	0.525&	0.473&0.523&0.385&	0.344&0.349\\
$i^{-2}$    &$U(-0.5,0.5)$   &200       &1000&	0.744&	0.640&0.759&0.646&	0.510&0.600\\
$i^{-2}$ 	&$U(-0.5,0.5)$ &	300 	&2000	&0.862	&0.774&0.880&0.815&	0.679&0.788\\
$i^{-2}$ 	&$U(-0.5,0.5)$&	400&	2000&	0.922&	0.770&0.917&0.871&	0.699&0.836\\
$i^{-2}$ 	&$U(-0.5,0.5)$&	1000&	4000&	0.997&	0.921&0.994&0.996&	0.880&0.985\\
$i^{-2}$    &$U(-0.5,0.5)$&	1500&	4000&	1    &	0.948 &0.999&	0.999&	0.932&0.999\\
$i^{-2}$ 	&$U(-0.5,0.5)$&	2000&	4000&	1    &	0.972&1&	1&	0.960&1\\
\hline
$i^{-2}$    &$U(-0.5,1)$	&100	&500&	1&	0.999&	1&1 &0.993&0.999\\
$i^{-2}$    &$U(-0.5,1)$&	200&	1000&	1&	1&	1&	1&1 &1\\
$i^{-2}$ 	&$U(-0.5,1)$&	300 	&2000	&1&	1&	1&	1&1 &1\\
$i^{-2}$ 	&$U(-0.5,1$ &	400&	2000&	1&	1&	1&	1&1 &1\\
$i^{-2}$ 	&$U(-0.5,1)$&	1000&	4000&	1&	1&	1&	1&1 &1\\
$i^{-2}$    &$U(-0.5,1)$&	1500&	4000&	1&	1&	1&	1&1 &1\\
$i^{-2}$ 	&$U(-0.5,1)$&	2000&	4000&	1&	1&	1&	1&1 &1\\
\hline
$i^{-2}$    &Sparse (a)	&100	&500&	\bf{1}&	0.045&0.139&\bf{0.951}&	0.011&0.044\\
$i^{-2}$    &Sparse (a)&	200&	1000&	\bf{1}&	0.036&0.147&\bf{0.952}&	0.013&0.038\\
$i^{-2}$ 	&Sparse (a)&	300	  &2000	&\bf{1}&	0.061&0.156&\bf{0.987}&	0.012&0.047\\
$i^{-2}$ 	&Sparse (a)&	400&	2000&	\bf{1}&	0.051&0.115&\bf{0.970}&	0.008&0.033\\
$i^{-2}$ 	&Sparse (a)&	1000&	4000&	\bf{1}&	0.065&0.121&\bf{0.952}&	0.012&0.036\\
$i^{-2}$    &Sparse (a)&	1500&	4000&	\bf{1}&	0.063&	0.073&\bf{0.899}&	0.014&0.012\\
$i^{-2}$ 	&Sparse (a)&	2000&	4000&	\bf{0.994}&	0.043&0.078&\bf{0.842}&	0.008&0.026\\
\hline
$i^{-2}$    &Sparse (b)	&100	&500&	0.904&	0.777&	0.970&0.647&0.400&0.736\\
$i^{-2}$    &Sparse (b)&	200&	1000&	0.904&	0.654&0.970&0.658&	0.368&0.742\\
$i^{-2}$ 	&Sparse (b)&	300	  &2000	& 0.970 &	0.632&	0.989&0.733&	0.357&0.829\\
$i^{-2}$ 	&Sparse (b)&	400&	2000&	0.907&	0.537&	0.959&0.647&	0.299&0.714\\
$i^{-2}$ 	&Sparse (b)&	1000&	4000&	0.864&	0.337& 0.932&0.579&	0.158&0.673\\
$i^{-2}$    &Sparse (b)&	1500&	4000&	0.654&	0.213& 0.830&0.424&	0.088&0.506\\
$i^{-2}$ 	&Sparse (b)&	2000&	4000&	0.555&	0.190&0.703&0.310&	0.074&0.414\\
\hline
\end{tabular}
\end{center}
\end{table}

By the Table \ref{tab2}, for the symmetric change of variance $U(-0.5,0.5)$, our proposed test $T_U$ and test $T_{LV}$ by \cite{Choi2021} outperformed the test $T_{V}$ by \cite{li2015}; for the asymmetric change of variance $U(-0.5,0.1)$, all the tests $T_U$, $T_V$ and $T_{LV}$ demonstrated outstanding performance; for the sparse change (a) of variance changes non-monotonically with positive and negative values, our test $T_{U}$ demonstrated superior performance, while tests $T_V$ and $T_{LV}$ showed the poor results with minimal power; for sparse change (b) of variance monotonically changes with negative values, $T_{LV}$ test performed best, followed by $T_U$ test, while $T_V$ test performed worst with minimal power. On the one hand, the absolute values of changes in sparse change (b) are smaller than that in sparse change (a), so the performances of test $T_U$ were not as good as that in sparse change (a). On the other hand, variance changes in practice are more likely to involve both positive and negative shifts. Therefore, our proposed test $T_U$ demonstrates overall superiority over tests $T_V$ and $T_{LV}$.

In addition, the performances of tests $T_U$, $T_V$ and $T_{LV}$ in Table 3 (see Appendix A), based on Gamma errors, are similar to those in Table \ref{tab2}, which are based on Gaussian errors. Furthermore, we consider the change-point location $t^*$ to be $\lfloor T/3\rfloor$, and obtain the empirical power and accuracy in the Tables 4 and 5 (see Appendix A). The results of Tables 4 and 5 are similar to those in Tables \ref{tab2} and 3. Here we did not give the details.

\setcounter{equation}{0}
\section{The Real Data Analysis}

In this section, we do the detection of the variance change-points based on the panel data of the Shanghai Shenzhen CSI 300 Index Components, which comes from \url{https://www.csindex.com.cn/#/indices/family/detail?indexCode=000300}.
Following the simulation section, $T_U$ denotes our test defined by \eqref{c6}, $T_V$ denotes the test of \cite{li2015} defined by \eqref{cr3}, while $T_{LV}$ denotes the test from \cite{Choi2021}. The data from 2006-01-01 to 2016-12-31 have 300 stocks, among which 4 stocks such as 302132, 688599, 688981, 001391 have missing values. So we consider the returns of Adjusted Closing Prices (ACP) for these no missing 296 stocks, where the ACP is calculated by multiplying the closing price by the adjustment factor. For example, consider the return
\begin{equation}
r_{i,t}=\log P_{i,t}-\log P_{i,t-1},\label{h1}
\end{equation}
where $P_{i,t}$ is the ACP of $i$-th stock at $t$ time, $P_{i,0}=0$, $1\leq i\leq 296$ and $1\leq t\leq 2673$.

Typically, after applying logarithmic transformation and differencing, changes in mean and variance become sparse.
Thus, we focus on the variance changes for the panel data of return $\{r_{i,t}\}$.
Employing binary method at a significance level $\alpha=0.05$, our proposed test $T_U$ (with $h_T=\lfloor T^{1/3}\rfloor$)
successfully detected 5 variance change-points: 2008-04-23, 2009-07-24, 2010-09-20, 2015-03-25, and 2015-10-26. In contrast,
the test $T_{LV}$ by \cite{Choi2021} detected only 2 change-points (2014-11-27 and 2016-03-02), where the critical value
was 40.1. But the test $T_V$ by \cite{li2015} failed to detect any change-points.

\begin{figure}[htbp]
\begin{center}\includegraphics[width=0.95\textwidth]{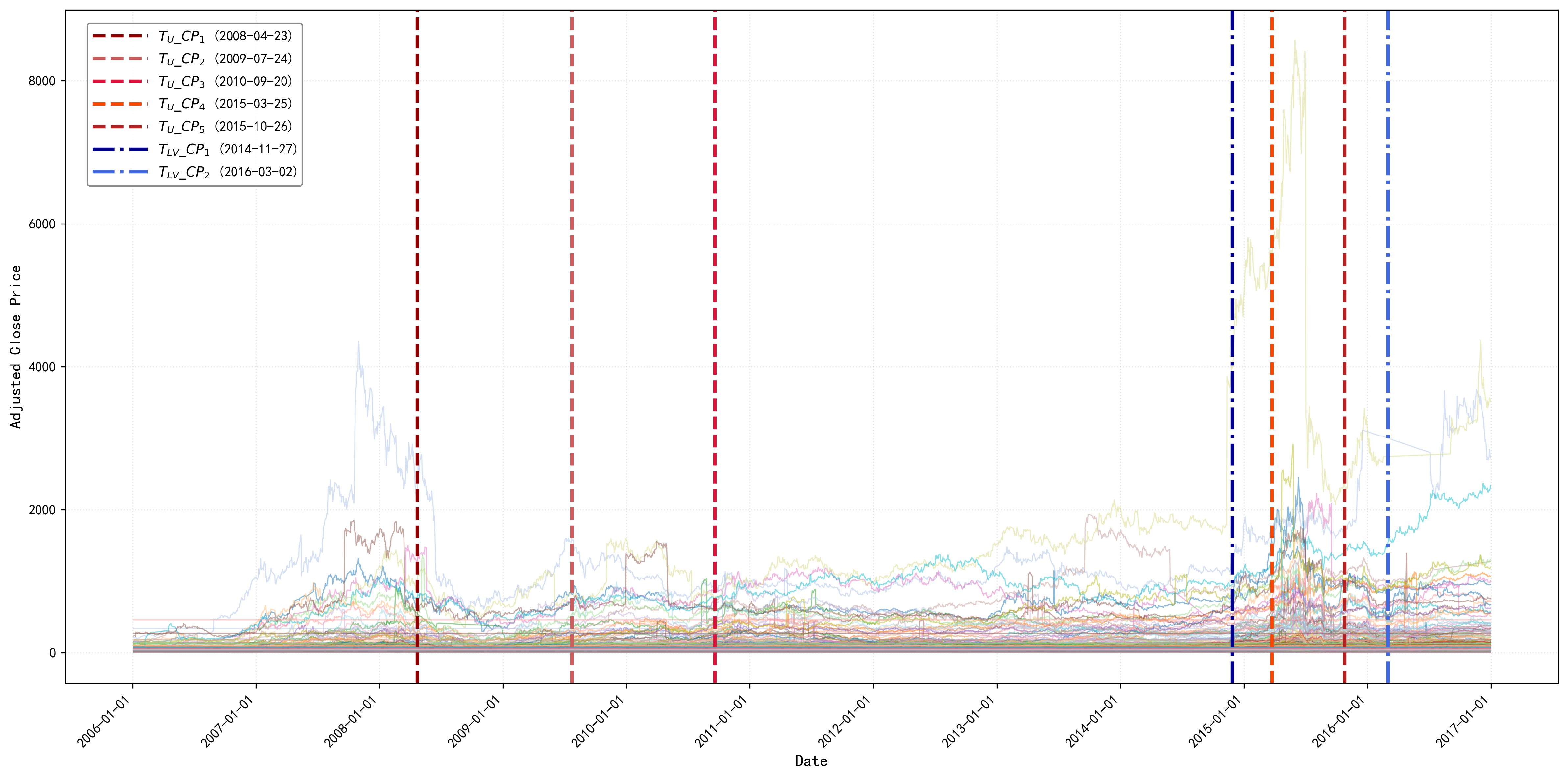}
\caption{The adjusted closing price chart and variance change-point of Shanghai Shenzhen CSI 300 index components, 296 stocks from 2006-01-01 to 2016-12-31}
\label{fig1}
\end{center}
\end{figure}

The change-points detected by tests $T_U$ and $T_{LV}$ are presented in the Figure \ref{fig1}.

\begin{figure}[htbp]
\begin{center}\includegraphics[width=0.95\textwidth]{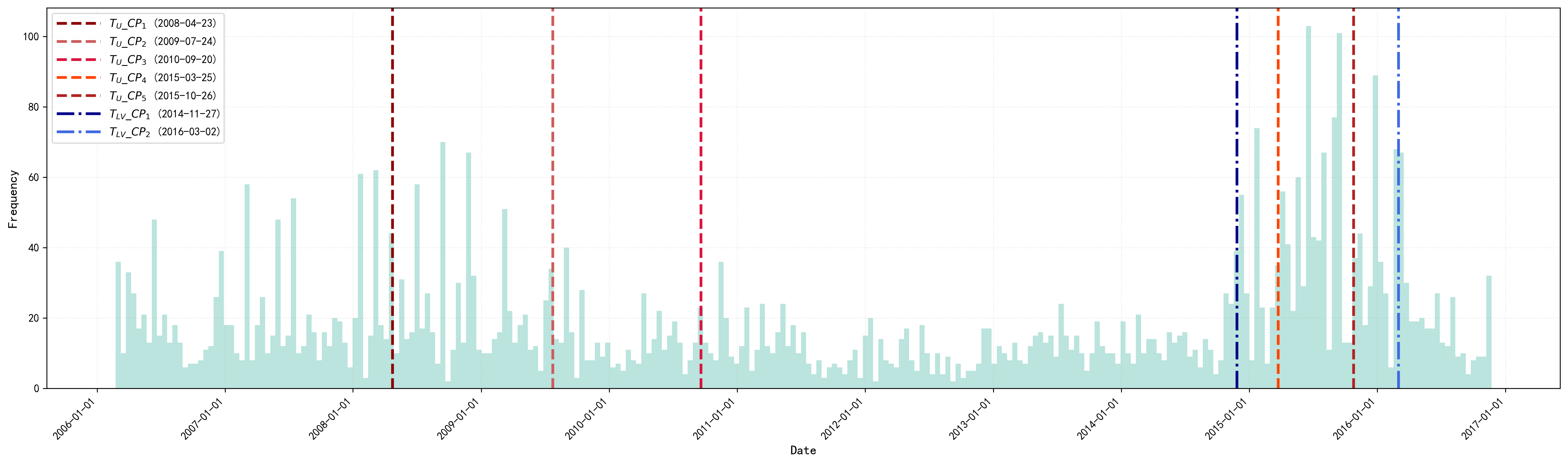}
\caption{The histogram of variance change-points with `cpt.var' method for 296 stocks from 2006-01-01 to 2016-12-31}
\label{fig2}
\end{center}
\end{figure}

In Figure \ref{fig1}, the notation $T_{U}\_CP\_1$(2008-04-23) denotes the first change-point detected by test $T_U$, which occurred on 2008-04-23. The remaining symbols are defined analogously. It can be observed that two change-points (2014-11-27 and 2016-03-02) detected by \cite{Choi2021}'s test $T_{LV}$ were in close to those detected by our proposed $T_U$ (2015-03-25, and 2015-10-26). However, our test detected three additional change-points (2008-04-23, 2009-07-24, 2010-09-20) that were missed by $T_{LV}$. Meanwhile, the test $T_V$ by \cite{li2015} failed to detect any change-points. On the other hand, for each return $r_{i,t}$ of $i$-th stock, we also use the binary method and the `cpt.var' method by the R Package \textbf{changepoint} of \cite{Killick2014} to detect the change-point of variances at the significance level $\alpha=0.05$. We give the histograms of these change-points in Figure \ref{fig2}.

As illustrated in Figure \ref{fig2}, the seven change-points detected by tests $T_U$ and $T_{LV}$
exhibit high frequencies. To verify the financial implications of the change-points, we conducted a correlation analysis between the detection results and key historical events in the A-share market, as well as the macroeconomic context.
\begin{itemize}
\item[(1)] Financial Crises and Policy Stimulus Cycles (2008-2010). During this period, the market was primarily influenced by exogenous macroeconomic shocks and countercyclical adjustment policies.
\begin{itemize}
\item[$\bullet$]  Performance of $T_{LV}$: No significant change-points were detected within this interval. At the significance level $\alpha=0.05$, the critical value 40.1 of $T_{LV}$ is relatively high. This implies that $T_{LV}$ exhibited a high rate of false negatives when volatility was driven purely by fundamental and policy factors, in the absence of leveraged fund amplification.
\item[$\bullet$]  Performance of $T_{U}$: Successfully identified three key structural turning points within this cycle:
\begin{itemize}
\item[$\circ$] 2008-04-23: Corresponding to the announcement of the stamp duty reduction policy (on the evening of April 23, 2008, the Ministry of Finance and the State Administration of Taxation announced that, starting April 24, the securities transaction stamp duty rate would be reduced from 3 \textperthousand~to 1\textperthousand ). The market ended its unilateral downward trend and entered a phase of high-volatility contention. $T_U$ accurately captured the restructuring of the volatility structure following the release of downside risks.
\item[$\circ$] 2009-07-24: Corresponding to the point of diminishing marginal returns of the ``four trillion yuan'' stimulus package. The State Council executive meeting proposed the measures on November 5, 2008, followed by the official announcement on the evening of November 9, unveiling ten measures to further boost domestic demand and promote economic growth, with a total scale of approximately four trillion yuan. This timing accurately marked the peak zone of the 2009 recovery rally (the Shanghai Composite Index peaked in early August), reflecting a shift in market expectations from policy easing to concerns about stimulus withdrawal, serving as a forward-looking indicator of an impending trend reversal.
\item[$\circ$] 2009-07-24: Corresponding to the start of the ``National Day Rally'' and rising inflation expectations. Market sentiment shifted from small and mid-cap thematic stocks to cyclical and heavyweight stocks, leading to a marked increase in the volatility midpoint.
\end{itemize}
\end{itemize}
\item[(2)] The Leverage Cycle and Abnormal Volatility (2014-2016). This phase exhibited typical capital-driven characteristics, as the market went through a complete cycle from valuation recovery to irrational exuberance, and ultimately to a liquidity crisis.
\begin{itemize}
\item[$\bullet$] Bull Market Initiation Phase:
\begin{itemize}
\item[$\circ$] $T_{LV}$ (2014-11-27): It accurately identified the first high-volatility window following the central bank's interest rate cut. On the evening of November 21, 2014, the People's Bank of China announced a reduction in the benchmark interest rates for RMB loans and deposits of financial institutions, effective from November 22. Thus, this change-point marked the official establishment of the ``leverage-driven bull market'' fueled by declining risk-free rates, as market volatility underwent a discontinuous surge, breaking free from a prolonged period of stagnation.
    \item[$\circ$] $T_U$ (2015-03-25): It identified the acceleration phase of the market rally. At this change-point, the Shanghai Composite Index broke through a key resistance level, and massive amounts of off-exchange margin financing entered the market. The signal from $T_U$ indicated a qualitative shift in market dynamics, transitioning from a rational valuation recovery to a speculative bubble, with a significant expansion of risk exposure.
\end{itemize}
\item[$\bullet$] Bubble Burst and Bottoming Phase:
\begin{itemize}
\item[$\circ$] $T_U$ (2015-10-26): Marked the end of the rebound following ``Stock Market Crash 2.0''. From August 18 to 26, 2015, after the initial crash in June, the A-share market experienced another cliff-like decline in mid-to-late August, particularly a consecutive plunge starting August 24 (``Black Monday''), with the Shanghai Composite Index falling below 3000 points within days. The term ``rebound'' here specifically refers to the recovery rally after Stock Market Crash 2.0 (the August Crash). As regulatory scrutiny intensified (e.g., the Xu Xiang case), market volatility retreated from a state of panic-driven highs to a gradual, grinding decline, signaling the final phase of the deleveraging process. On November 1, 2015, Xu Xiang was apprehended on the Hangzhou Bay Bridge in Ningbo, leading to a swift retreat of speculative capital and fundamentally altering the nature of the rebound.
    The change-point on October 26 (a few days before Xu Xiang's arrest) indicates that the algorithm astutely detected the preemptive sell-off by major funds ahead of the risk event.
\item[$\circ$] $T_{LV}$ (2016-03-02): Marked the absolute low point after the ``circuit breaker'' crisis. Following the extreme liquidity squeeze in early 2016, market volatility contracted sharply, subsequently entering a prolonged phase of low-volatility, slow-growth uptrend. The A-share market circuit breaker mechanism was officially implemented on January 1, 2016. It was triggered twice-on January 4 (the first trading day) and January 7, leading to early market closures. On the evening of January 7, the China Securities Regulatory Commission announced the suspension of the mechanism. After hitting a low of 2638 points (January 27, 2016), the market underwent a month of repeated bottoming consolidation. In early March 2016 (particularly around the ``Two Sessions'' held on March 5 and March 3, 2016), the market established a pattern of not breaking below previous lows and launched the well-known ``2016-2017 blue-chip slow bull'' rally.
\end{itemize}
\end{itemize}
\end{itemize}

The empirical results indicate that the comparison method $T_{LV}$ functions more as a ``crisis early-warning system''. It demonstrates a high signal-to-noise ratio in capturing extreme market risks (such as the post-circuit breaker bottom in early 2016), making it well-suited for strategic timing decisions. In contrast, our method $T_U$ proposed in this paper serves as a ``market structure monitor for all seasons''. It exhibits superior sensitivity and interpretability in handling macroeconomic cycles not driven by leverage (2008-2010) as well as distinct market phases (such as the bubble formation in 2015 and the protracted decline in 2016). Our $T_U$ test is capable of delineating the evolutionary path of market volatility in greater detail, thereby providing richer informational dimensions to support medium to long-term macro risk management and asset allocation.

\setcounter{equation}{0}
\section{Conclusion and Discussion}
On the one hand, the CUSUM method is a widely used technique for change-point analysis. On the other hand, panel data models play a crucial role in economics and finance. \cite{horvan2012} investigated mean change-point detection in panel data models with errors following a linear process based on $i.i.d.$ random variables. \cite{li2015} extended this framework to variance change-point detection. Given that $\alpha$-mixing is a reasonably weak dependence condition with broad applications in time series analysis, this paper further examines variance change-points in the panel data model \eqref{a1} with $\alpha$-mixing errors. Under the null hypothesis of no variance change, Theorem 3.1 establishes the limit distribution of the CUSUM statistic. To estimate the long-run variances, Lemma 3.1 provides the consistency rates of the corresponding estimators. Consequently, Theorem 3.2 presents the CUSUM test $T_U$ of variance change-point detection, which can be effectively used to identify structural breaks in variance. We also also analyze the asymptotic behavior of the CUSUM statistic under alternative hypothesis. To evaluate the performance of our test, we conduct simulations for panel data models with Gaussian and Gamma errors. For comparison, we give the empirical size, power and accuracy of our test $T_U$, \cite{li2015}'s test $T_V$ and \cite{Choi2021}'s test $T_{LV}$ under symmetric, asymmetric and sparse variance change.
In the case of symmetric variance changes, tests $T_U$ and test $T_{LV}$ outperformed the test $T_{V}$. For asymmetric changes, all the tests $T_U$, $T_V$ and $T_{LV}$ demonstrated excellent performance. For the sparse change (a), where variance changes non-monotonically with both positive and negative values, test $T_{U}$ achieved superior performance, while tests $T_V$ and $T_{LV}$ exhibited poor results with minimal power. For the sparse change (b), characterized by monotonically decreasing variance, $T_{LV}$ test performed best, followed by $T_U$ test, while $T_V$ test performed worst with minimal power. It is worth noting that the absolute magnitude of variance changes in sparse change (b) is smaller than in sparse change (a), which explains why the performance of $T_U$ in case (b) was not as good as in case (a).
Moreover, in practice, variance changes are more likely to involve both positive and negative shifts. Therefore, overall, our proposed test $T_U$ demonstrates superior performance compared to tests $T_V$ and $T_{LV}$. Finally, we apply our method to a real data set from Shanghai Shenzhen CSI 300 Index Components, identifying 5 variance change-points and providing economic interpretations for the observed volatility shifts. In contrast, \cite{Choi2021}'s test $T_{LV}$ detected only 2 change-points, while the test $T_V$ by \cite{li2015} failed to detect any change-points. For future research, it would be interesting to explore change-point detection in panel data models when the number of panels $N$ exceeds the number of time periods $T$.

\vspace{0.5cm}
\noindent \textbf{Disclose Statement}

The authors have no conflict of interest to disclose.

\noindent \textbf{Funding}

This research is partially supported by NSSFC grant 24BTJ073.

\bibliography{bibliography.bib}

\end{document}